# MODERN TOOLS FOR ANNOTATION OF SMALL GENOMES OF NON-MODEL EUKARYOTES

M.A. Galchenkova, A.A. Korzhenkov*

NRC Kurchatov Institute, Kurchatova sq. 1, 123182, Moscow, Russia

* e-mail: Korzhenkov_AA@nrcki.ru

**Summary**. Nowadays, due to the increasing amount of experimental data obtained by sequencing, the most interest is focused on determining the functions and characteristics of its individual parts of the genome instead of determining the nucleotide sequence of the genome. The genome annotation includes the identification of coding and non-coding sequences, determining the structure of the gene and determining the functions of these sequences.

Despite the significant achievements in computational technologies working with sequencing data, there is no general approach to the functional annotation of the genome in the reason of the large number of unresolved molecular determination of the function of some genomes parts. Nevertheless, the scientific community is trying to solve this problem. This review analyzed existing approaches to eukaryotic genome annotation.

This work includes 3 main parts: introduction, main body and discussion. The introduction reflects the development of independent tools and automatic pipelines for annotation of eukaryotic genomes, which are associated with existing achievements in annotating prokaryotic ones. The main body consists of two distinguished parts, the first one is devoted to instructions for annotating genomes of non-model eukaryotes, and the second block is about recent versions of automatic pipelines that require minimal user's curation. The question of assessing the quality and completeness of the annotated genome is noted briefly, and the tools to conduct this analysis are discussed.

Currently, there is no universal automatic software for eukaryotic genome annotation, covering the whole list of tasks, without manual curation or using additional external tools and resources. Thus it leads to the task of developing a wider functional and universal protocol for automatic annotation of small eukaryotic genomes.

# СОВРЕМЕННЫЕ ИНСТРУМЕНТЫ АННОТАЦИИ МАЛЫХ ГЕНОМОВ НЕМОДЕЛЬНЫХ ЭУКАРИОТ

М.А. Галченкова, А.А. Корженков*

НИЦ Курчатовский институт, пл. Академика Курчатова, д. 1, 123182 Москва, Россия

* e-mail: Korzhenkov_AA@nrcki.ru

**Аннотация.** На сегодняшний день в связи с возрастающим количеством полученных экспериментальных данных секвенирования растет потребность не столько к определению нуклеотидной геномной последовательности, сколько к поиску и описанию функциональных единиц генома. Аннотация генома включает в себя идентификацию кодирующих и некодирующих последовательностей, определение структуры гена, включая нетранслируемые регионы, интроны и экзоны, и определение функций найденных последовательностей. Несмотря на имеющиеся значимые достижения в вычислительных технологиях, направленных на работу с геномными и транскриптомными данными, обработка эукариотических геномов является трудной вычислительной задачей, а разнообразие инструментов может поставить исследователя, впервые сталкивающегося с такой задачей в затруднительное положение.

В данной работе был выполнен обзор существующих подходов к аннотации геномов эукариот, включающий как отдельные инструменты, так и полуавтоматические программные конвейеры для аннотации геномов эукариот. В первой части обзора детально рассмотрен пошаговый алгоритм аннотации эукариотических геномов с указанием программного обеспечения и баз данных, применяемых для решения данных задач. Вторая часть посвящена существующим актуальным автоматизированным программным конвейерам для аннотации эукариотических геномов. Дополнительно, упомянуты инструменты для оценки качества и полноты генома.

Несмотря на наличие разработанного подхода к аннотированию геномов эукариот, не существует универсального автоматического программного обеспечения, способного выполнить весь спектр поставленных задач. Современные программные конвейеры имеют ряд недостатков, включая многочисленные зависимости, сложность настройки и запуска, низкую универсальность, невозможность автономной работы, что подчеркивает необходимость разработки программного обеспечения для автоматической аннотации малых геномов немодельных эукариот.

**Ключевые слова:** биоинформатика, геномика эукариот, структурная аннотация, функциональная аннотация

**Введение**

Повышение доступности высокопроизводительного секвенирования вызвало экспоненциальный рост объема геномных данных. На примере базы NCBI RefSeq, которая стремится к максимальному сокращению избыточности данных, хорошо видно, что трехкратный рост обусловлен не просто повышением количества данных, но и повышением их разнообразия — качественным ростом. Так, база данных **NCBI RefSeq** за 5 лет от версии 74 (11 января 2016, https://ftp.ncbi.nlm.nih.gov/refseq/release/release-notes/archive/RefSeq-release74.txt) к версии 204 (4 января 2021, https://ftp.ncbi.nlm.nih.gov/refseq/release/release-notes/RefSeq-release204.txt) выросла с 22,36 млрд аминокислотных остатков (а.о.) в 58,5 млн последовательностей до 73.97 млрд а.о. (331%) в 191,41 млн последовательностей (327%). В тоже время размер баз **NCBI GenBank** и **NCBI WGS** в течение 5 лет вырос в 8,4 раз: с 1,5 трлн до 12,55 трлн (декабрь 2020) оснований (https://www.ncbi.nlm.nih.gov/genbank/statistics/). Такой значительный рост объема данных требует разработки производительных и надежных инструментов для обработки, анализа и хранения данных высокопроизводительного секвенирования, включая собранные геномные последовательности, в частности, автоматизированных инструментов для конвейерной обработки данных как для повышения воспроизводимости результатов, так и для повышения производительности труда научных работников.

Разработка средств автоматизированной аннотации геномов прокариот привела к появлению множества онлайн и оффлайн инструментов, запрашивающих у пользователя только исходную геномную последовательность и предоставляющих результат аннотации в виде набора файлов в популярных форматах (GFF, GTF, GenBank, FASTA и др.). Онлайн инструменты могут функционировать отдельно, например, **RAST** (https://rast.nmpdr.org/; Brettin et al., 2015), **KBase** (https://kbase.us/; Arkin et al. 2018), **IGS Prokaryotic Annotation Pipeline** (https://www.igs.umaryland.edu/services/analysis.php; Galens et al., 2011) или быть интегрированы в процесс депонирования геномов в базы данных как **NCBI PGAP** (Tatusova et al., 2016), **IMG M/ER** (Huntemann et al., 2015).

Очевидными достоинством этих систем является графический интерфейс, отсутсвие установки, необходимости скачивания баз данных, обеспечение вычислительных ресурсов и пр., однако, стоит помнить, что такие инструменты не могут быть использованы для анализа конфиденциальных данных. В качестве оффлайн инструментов, предоставляющих полный спектр инструментов анализа и форматирование полученных данных в соответствии со стандартами INSDC для последующей публикации и дружественных в процессе установки и использования, можно отметить **Prokka** (Seemann, 2014) и **DFAST** (Tanizawa et al., 2018).

Ввиду большего размера геномов эукариот и более сложного механизма экспрессии генов, поточный анализ эукариотических геномов представляется более сложной задачей. Для ряда модельных организмов существуют специализированные базы данных, содержащие аннотированные референсные геномы. В первую очередь стоит отметить портал **Ensembl** (https://www.ensembl.org), на котором на середину 2020 года были представлены геномные данные 238 позвоночных, 111 прочих животных, 71 растения, 237 протистов, 1014 грибов, а также доступен ряд инструментов для работы с этими данными (Yates et al., 2020). Несмотря на широкий охват эукариот, стоит отметить, что при анализе немодельных организмов, эти данные могут быть использованы для сравнительного анализа, но непосредственный перенос аннотаций может привести к серьезным ошибкам.

В данном обзоре авторы представляют современные подходы и инструменты функциональной аннотации малых геномов эукариот. Необходимо подчеркнуть, что в данном описании авторы не проводят направленного сравнения имеющихся подходов, т.к., во-первых, для ряда инструментов существуют ограничения на спектр анализируемых геномов и требования к входным данным, во-вторых, они включают в себя разнообразные и частично перекрывающиеся наборы отдельных инструментов, которые влияют на конечный результат и могут быть взаимозаменяемы с аналогами в зависимости от параметров запуска. Также стоит отметить что поддержка и доступность существующих инструментов может быть прекращена, а новые биоинформатические инструменты появляются без преувеличения ежедневно, в связи с чем ни один обзор не может претендовать на полное и исчерпывающее описание.

**Основные этапы аннотации эукариотических геномов**

Аннотация малых геномов эукариот включает в себя три основных этапа: определение и маскировка повторяющихся последовательностей, структурная аннотация и функциональная аннотация (Yandell, 2012). В качестве альтернативы выделяют идентификацию некодирующих участков генома, поиск открытых рамок считывания и функциональную аннотацию белок-кодирующих генов (Haridas S. et al., 2018). По сравнению с прокариотическими организмами геномы эукариот содержат значительное количество повторяющихся последовательностей, которые усложняют процесс аннотирования и должны быть предварительно выявлены в анализируемом геноме. К структурной аннотации относят задачи идентификации белок-кодирующих последовательностей, включая определение сайтов сплайсинга, генов функциональных РНК (тРНК, рРНК, микроРНК и др.). Вслед за структурной аннотацией следует функциональная, которая заключается в определении функциональных признаков аннотированных структур. Существует большое разнообразие программных инструментов для аннотирования геномов, различающихся по условиям использования: свободное, бесплатные академические лицензии, исключительно платное использование, открытости и лицензии исходного кода, доступности онлайн и возможности локальной установки. В обзоре мы по возможности сосредоточим внимание на инструментах с открытым исходным кодом или как минимум доступных бесплатно для академического использования. Инструменты и базы данных выделены **полужирным начертанием** и при первом упоминании указывается ссылка на статью, а также, при наличии веб-интерфейса, поддерживаемого разработчиками, приводится ссылка и на него.

*Идентификация некодирующих участков генома*

- поиск и маскирование повторов и транспозонов: **RepeatModeler** (http://www.repeatmasker.org/RepeatModeler/), **RepeatModeler2** (Flynn et al., 2020),

**RepeatMasker** (Tempel, 2012), **tantan** (Frith, 2011), база данных **Dfam** (Hubley et al., 2016);

- поиск генов тРНК: **tRNAscan-SE** (http://lowelab.ucsc.edu/tRNAscan-SE/; Lowe et al., 1997; Chan et al, 2019), **Aragorn** (Laslett & Canbäck, 2004), **Arwen** (Laslett & Canbäck, 2008);

- идентификация генов малых ядрышковых РНК — **snoSeeker** (Yang J. H. et al., 2006), генов микроРНК — **miRDeep** (An J. et al., 2013), **miRanalyzer** (Hackenberg et al., 2011); генов прочих нкРНК — **ERPIN** (Gautheret et al., 2001), **miRAlign** (Wang X. et al., 2005);

- альтернативно, поиск генов некодирующих РНК можно осуществить при помощи инструментов cmsearch и cmscan из пакета **Infernal** (Eddy et al., 2013) и ковариационных моделей базы данных **Rfam** (https://rfam.xfam.org/; Kalvari et al., 2018).

*Определение белок-кодирующих генов*

Определение белок-кодирующих генов возможно при помощи одного из следующих подходов, либо их комбинации:

1. Предсказание генов *ab initio* может быть проведено при помощи **GeneMark-ES** (Ter-Hovhannisyan et al., 2008), **FGENESH** (Zhang S. et al., 2008), **AUGUSTUS** (Keller et al., 2011), **SNAP** (Korf, 2004) и **GlimmerHMM** (Majoros et al., 2004).
2. Предсказание генов на основе гомологии осуществляется путем выравнивания последовательностей из баз данных, например, **UniProt/SwissProt** (https://www.uniprot.org/) против анализируемого генома при помощи **GeneWise** (Birney et al., 2004), **Spaln** (Gotoh, 2008), **exonerate** (Slater & Birney, 2005), **GeneMark-EP+** (Brůna et al., 2020) или альтернативных программ.
3. Предсказание генов, основанное на транскриптомных данных включает сборку транскриптов *de novo* или на основании картирования прочтений и последующее выравнивание транскриптов на геном.

a. *de novo* сборка транскриптов может быть осуществлена при помощи **Trinity** (Grabherr et al., 2011), **rnaSPAdes** (Bushmanova et al., 2019), **Trans-ABySS** (Robertson et al., 2010), **Bridger** (Chang Z., 2015) и **SOAPdenovo-Trans** (Xie Y. et al., 2014), перечисленные программы показали лучшие результаты в *de novo* сборке транскриптома по результатам сравнительного анализа (Hölzer & Marz, 2019).

b. Сборка транскриптов на основе картирования прочтений, их выравнивание транскриптов и предсказание генов может быть проведено при помощи **TopHat** и **Cufflinks** (Trapnell et al., 2012), **STAR** (Dobin et al., 2013), **PASA** (Haas et al., 2003), **GeneMark-ET** (Lomsadze et al., 2014) или **mGene.ngs** (Behr et al., 2010).

Оценка и сопоставление предсказаний могут быть проведены при помощи средств визуализации, в частности **IGV** (https://igv.org/; Thorvaldsdóttir et al., 2013). Для оценки, фильтрации и консолидации результатов предсказания геном несколькими инструментами могут быть использованы **EvidenceModeler** (Haas et al., 2008), **MAKER** (Cantarel B. L. et al., 2008) и другие программы, рассматриваемые во второй части обзора.

*Функциональная аннотация*

Функциональной аннотации проводится на основании предсказанных белок-кодирующих последовательностей и включает в себя 3 общих подхода:

(1) характеристика частей последовательности белка, таких как домены;

(2) поиск гомологии с уже охарактеризованными последовательностями;

(3) аннотация в соответствии с существующими схемами классификации: **Eu*k*aryotic *O*rthologous *G*roups** (**KOG**; Koonin E. V. et al., 2004), **Gene Ontology** (**GO**; http://www.geneontology.org; Ashburner et al., 2000), **Kyoto Encyclopedia of Genes and Genomes** (**KEGG**; http://www.genome.jp/kegg; Kanehisa et al., 2017), **Enzyme Commission** (**EC**, http://www.expasy.org/enzyme/) и др.

Функциональная аннотация белок-кодирующих генов может включать следующие этапы:

a. Определение доменов из базы **Pfam** (http://pfam.xfam.org/; El-Gebali et al., 2019) при помощи **hmmscan** (http://hmmer.org; Eddy, 2011).

b. Идентификации сигнальных пептидов, предполагающих секрецию белка в предсказанных генах **signalP** (http://www.cbs.dtu.dk/services/SignalP/; Almagro Armenteros et al., 2019).

c. Предсказание трансмембранных доменов при помощи **TMHMM** (http://www.cbs.dtu.dk/services/TMHMM/; Krogh er al., 2001).

d. Функциональный анализ белков, включающий предсказание доменов на основании сигнатур, полученных при анализе различных баз данных, и классификацию последовательностей при помощи **InterProScan** (Quevillon E. et al., 2005; http://www.ebi.ac.uk/interpro/download.html).

e. Парное выравнивание последовательности гена с помощью **NCBI BLAST** (Camacho et al., 2009), **BLAT** (Kent, 2002), **diamond** (Buchfink et al., 2015) или аналогичных инструментов против универсальных баз, таких как **NCBI nr** (https://www.ncbi.nlm.nih.gov/nuccore/, https://www.ncbi.nlm.nih.gov/protein/), **SwissProt** (http://www.expasy.org/sprot/) или **UniProt** (https://www.uniprot.org/) или специализированных баз, например **MEROPS** (http://merops.sanger.ac.uk; Rawlings et al., 2010) для идентификации пептидаз.

f. Классификация генов:

(1) Присвоение терминов **GO** одной из трех категорий: биологический процесс, молекулярная функция и клеточный компартмент при помощи баз данных **Interpro** (http://www.ebi.ac.uk/interpro/; Finn et al., 2017) и **SwissProt**.

(2) Сопоставление генов с ферментами метаболических путей базы **KEGG** и присвоение кодов **EC** при помощи **KEGG Mapper** (Kanehisa & Sato, 2020).

(3) Отнесение к кластеру ортологичных эукариотических генов **KOG**.

g. Поиск групп ортологичных генов при помощи: **OrthoFinder** (Emms et al., 2015), **OrthoMCL** (Li L. et al., 2003).

*Аннотация генома митохондрий*

Отдельно стоит выделить аннотацию митохондриального генома, т.к. генетический материал митохондрий отличается от ядерного и требует использования дополнительных инструментов. Аннотация митохондриального генома состоит из следующих основных шагов:

1. Предсказание генов, в том числе фрагментированных, кодирующих тРНК, с помощью **tRNAscan-SE** (Chan P. P. et al, 2019), **ARWEN** (Laslett D. et al., 2008) и **RNAweasel** (Gautheret D. et al., 2001).
2. Предсказание белок-кодирующих генов возможно по одному из следующих методов:
    a. *de novo* предсказание с использованием генетического кода и моделей генов, специфичных для митохондриальных геномов;
    b. Выравнивание интрон-содержащих генов с использованием **TBLASTN** (Gertz et al., 2006) без учета согласований сайтов сплайсинга и уточнение границ с сохранением рамки считывания или с использованием **GeneWise** (Birney et al., 2004).
3. Предсказание генов рРНК с помощью **Infernal** (Eddy et al., 2013) с использованием ковариационных моделей генов малой и большой субъединиц рРНК из базы данных **Rfam** (Kalvari et al., 2018).
4. Определение артефактов, вызванных линейным представлением сборки кольцевого генома (характерного для митохондрий животных), например повторяющихся фрагментов на обоих концах последовательности, отсутствующих в реальном митохондриальном геноме. Дупликации или фрагментация генов из канонического набора потенциально является признаком описанного артефакта сборки. В таких случаях рекомендуется ручная правка последовательности и ее повторная аннотация.

Вышеописанные подходы весьма специфичны для каждого набора данных и требует непосредственного включения пользователя на этапе очистки с подготовкой для дальнейшей аннотации и фильтрации полученных предсказанных функциональных участков генома, что увеличивает время работы с данными. Также отсутствие единого автоматизированного протокола существенно усложняет процесс постобработки экспериментальных данных.

**Автоматические системы аннотации эукариотических геномов**

Данный раздел обзора посвящен реализованным (полу)автоматическим пайплайнам, комбинирующим в себе различные внешние инструменты, упомянающиеся ранее. Нижеописанные пайплайны аннотирования используют комбинацию предсказаний *ab initio* и предсказаний, основанных на фактических данных, для создания точных консенсусных аннотаций.

**MAKER** ([http://www.yandell-lab.org/software/maker.html](http://www.yandell-lab.org/software/maker.html)) — это полностью автоматизированный инструмент аннотации, который широко используется для аннотирования эукариотических геномов (Cantarel B. L. et al., 2008) и включает следующие этапы:

- поиск и маскирование повторов: **RepeatMasker** и **NCBI BLASTX** (Camacho et al., 2009),
- *de novo* предсказание белок-кодирующих генов: **GeneMark-ES** (Ter-Hovhannisyan et al., 2008), **FGENESH** (Zhang S. et al, 2008), **AUGUSTUS** (Keller et al., 2011), **SNAP** (Korf, 2004),
- выравнивание нуклеотидных и аминокислотных последовательностей **NCBI BLAST** (Camacho et al., 2009) с последующей оптимизацией выравнивания (**exonerate** (Slater & Birney, 2005)).

Полученные данные подвергаются автоматическому анализу для выбора оптимального предсказания гена, включая нетранслируемые регионы и сайты альтернативного сплайсинга, и проводится количественная оценка качества предсказания. Для тестирования инструмента работает онлайн-версия — [http://weatherby.genetics.utah.edu/cgi-bin/mwas/maker.cgi](http://weatherby.genetics.utah.edu/cgi-bin/mwas/maker.cgi) с ограничениями на длину аннотируемых последовательностей. MAKER бесплатно доступен для научной работы, однако необходима предварительная регистрация.

**BRAKER1** (Hoff K. J. et al., 2016) и его расширение **BRAKER2** (Brůna et al., 2021) — это инструмент для структурной аннотации геномов с возможностью анализа транскриптомных данных и последовательностей гомологичных белков. **BRAKER** использует **GeneMark-ES/ET/EP** (Brůna et al., 2020) и **AUGUSTUS**. На первом этапе производится предварительное предсказание генов: **GeneMark-ES** проводит *ab initio* предсказание генов основываясь на геномной последовательности, при наличии транскриптомных данных используется **GeneMark-ET** (Lomsadze et al., 2014), при наличии гомологичных аминокислотных последовательностей генов, например из базы **OrthoDB**, они выравниваются на геном при помощи **ProtHint** (https://github.com/gatech-genemark/ProtHint) и на основании выравнивания **GeneMark-EP** проводит предсказание генов. Полученные предсказания используются на втором этапе для обучения инструмента **AUGUSTUS** и финального предсказания генов. **BRAKER** реализован на языке Perl и активно развивается.

**CodingQuarry** — это программный пакет, реализованный на языке C++, для предсказания генов на основании транскриптомных данных, оптимизированный для анализа геномов грибов, которые обладают меньшим, чем у высших эукариот, размером интронов (Testa et al., 2015). В этом инструменте транскриптом, собранный при помощи **Cufflinks** (Trapnell et al., 2012), используется для построения обобщенной скрытой марковской модели, которая используется для предсказания генов. Последнее обновление инструмента (версия 2.0) было опубликовано в 2016 году, что указывает на прекращение разработки.

**OMIGA** (Optimized Maker-Based Insect Genome Annotation) — это инструмент для аннотации геномов насекомых (Liu J. et al., 2014), основанный на программе **MAKER**. Кратко обработка данных состоит в маскировании повторов при помощи **RepeatMasker** и **RepeatModeler**. Затем проводится картирование транскриптомных прочтений на геном при помощи **Bowtie2** (Langmead & Salzberg, 2012) для определения транскрибируемых регионов и сборки транскриптов при помощи **Cufflinks**. Высококачественные транскрипты были использованы для обучения и последующего предсказания генов при помощи **AUGUSTUS**, **SNAP** и **GeneMark**. Структура генов была проверена и исправлена при помощи Exonerate. В завершение была проведена интеграция всех предсказаний при помощи **MAKER**. Несмотря на успешные результаты использования этого инструмента на реальных данных в настоящее время он недоступен для скачивания.

**EuGene** — это автоматизированный инструмент для поиска генов в прокариотических (Sallet et al., 2014) и эукариотических геномах (Sallet et al., 2019). EuGene использует:

- Инструменты для поиска геномных повторов: **Red** (Girgis, 2015), **LTRHarvest** (Ellinghaus et al., 2008), **NCBI BLASTX** и базу данных **Repbase Update** (Bao et al., 2015).
- Вероятностные модели межгенных, интронных, транскрибируемых и транслируемых регионов и их границ (например, сайты сплайсинга).
- Транскриптомные данные: картированные при помощи **GMAP** (Wu & Watanabe, 2005) прочтения или транскриптомные сборки.
- Инструменты выравнивания аминокислотных последовательностей **diamond** и **NCBI BLAST+** для поиска гомологов.
- Инструменты поиска функциональных РНК: **RNAmmer** или **Infernal**.
- Импортированную информацию специализированных инструментов в формате GFF3.

**EuGene** реализован на языке Perl, документация и исходный код доступны на веб-сайте http://eugene.toulouse.inra.fr/.

**GeMoMa** (Gene Model Mapper) позволяет проводить предсказание белок-кодирующих генов на основе геномных и, факультативно, транскриптомных данных (Keilwagen et al., 2016; Keilwagen et al., 2019). **GeMoMa** написан на языке Java и доступен в виде JAR-файла, не требующего установки. Аннотация включает несколько стадий:

- Обработка транскриптомных данных. Прочтения картируют на геном и определяют позиции интронов при помощи **TopHat2** или **STAR**.
- Поиск фрагментов белок-кодирующих генов. При помощи аннотированных сборок референсных геномов получают информацию о фрагментах белок кодирующих генов, отвечающих экзонам, комбинируемую с информацией об интронах с предыдущего шага для создания моделей генов.
- Поиск гомологичных последовательностей белок-кодирующих генов с помощью **TBLASTN** на основании референсных последовательностей.
- Предсказание транскриптов по результатам выравнивания. Предсказание сайтов сплайсинга проводится по данным транскриптомного секвенирования, либо при низком покрытии по консервативным сайтам GT/GC и AT. При наличии нескольких альтернативных вариантов экзонов в соответствующем регионе генома все комбинации, соответствующие рамке считывания рассматриваются и проходят оценку.
- На заключительном этапе проводится фильтрация предсказанных транскриптов по разным критериям, включая относительный балл **GeMoMa** для транскрипта, полноту транскрипта (наличие старт-кодона и стоп-кодона) и число референсных геномов, подтверждающих предсказанный транскрипт. Выходные данные — отфильтрованная и комбинированная аннотация в формате GFF.

Стоит отметить, что в отсутствие транскриптомных данных возможно использование только референсных геномных сборок. К недостаткам инструмента можно отнести невозможность предсказания нетранслируемых регионов гена (UTR).

**FunGAP** — это автоматизированный программный пакет для предсказания и аннотации белок-кодирующих генов в геномах грибов (Min B. et al., 2017). **FunGAP** использует три программы для предсказания белок-кодирующих генов: **AUGUSTUS**, **Braker** и **MAKER**. В качестве входных данных используется геномная последовательность, мРНК-прочтения и база референсных

аминокислотных последовательностей для поиска гомологичных генов. Аннотация состоит из трех этапов:

1. предварительная обработка: маскирование повторов в геноме и сборка транскриптома;
2. предсказание генов при помощи доступных входных данных тремя инструментами;
3. оценка предварительных результатов:
   - Выравнивание против референсного протеома с помощью **BLASTP** (Camacho et al., 2009).
   - Выравнивание против моделей консервативных однокопийных генов из базы **BUSCO** (Seppey et al., 2019).
   - Идентификация доменов **Pfam** при помощи **InterProScan**.

Каждый метод предоставляет количественную оценку, рассчитываемую как вес выравнивания умноженный на долю длины гена, покрытого этим выравниванием. Для каждого предсказанного гена эти значения суммируются и учитываются на этапе фильтрации как качество предсказания. В процессе фильтрации **FunGAP** находит «генные блоки» — наборы перекрывающихся генов — и для каждого блока отбирает предсказание с наивысшим качеством. Конечными данными являются: аминокислотные последовательности белков в формате FASTA, файл аннотации в формате GFF3 и общая сводка результатов в формате HTML. **FunGAP** реализован на интерпретируемом языке Python версии 2.7, что в свою очередь ограничивает интеграцию и поддержку этого инструмента.

**Funannotate** является программным пакетом широкого назначения для предсказания и аннотации генов и проведения сравнительного анализа эукариотических геномов (Palmer & Stajich, 2020). Изначально **funannotate** был предназначен для аннотирования геномов грибов и имел ограничение на размер генома до 30 млн п. н., но сейчас предоставляется возможность работы с геномами высших эукариот. Для аннотированных геномов можно провести сравнительный анализ, результат которого будет представлен в формате HTML.

**Funannotate** включает стадию предварительной обработки генома: удаление небольших повторяющихся контигов из сборки, сортировка и переименование заголовков контигов для совместимости с биоинформатическими инструментами и базами данных, маскирование повторов при помощи **tantan**, **RepeatMasker** или **RepeatModeler**.

Предсказание белок-кодирующих генов проводится при помощи **Evidence Modeler** (Haas et al., 2008), который использует данные различных инструментов для предсказания генов: **AUGUSTUS**, **SNAP**, **glimmerHMM**, **CodingQuarry** и **GeneMark-ES/ET**. В анализ кроме геномной последовательности могут быть дополнительно включены транскриптомные данные секвенирования и/или референсные последовательности генов и геномов. При наличии транскриптомных данных возможно уточнение предсказания генов и аннотации нетранслируемых областей при помощи **PASA** (Haas et al., 2008).

После предсказания белок-кодирующих генов проводится их функциональная аннотация: идентификация доменов **Pfam**, семейств **CAZYmes** (http://www.cazy.org/; Lombard et al., 2014), детекция сигнальных пептидов секретируемых белков при помощи **SignalP** (Almagro Armenteros et al., 2019), определение семейств протеаз базы **MEROPS** и принадлежность к группам ортологичных генов **BUSCO**. Дополнительно может быть проведена аннотация с помощью **InterProScan5**, которая включает присвоение терминов InterPro, онтологии GO и поиск транскрипционных факторов. Если **eggNOG-mapper** (Huerta-Cepas J. et al, 2017) установлен локально, то аннотации **eggNOG** и **COG** также могут быть добавлены к функциональной аннотации. Предсказание генов, отвечающих за вторичные метаболиты может быть проведено с использованием **antiSMASH** (https://antismash.secondarymetabolites.org/; Blin et al., 2019).

Результатом работы **funannotate** является набор файлов популярных биоинформатических форматов, в том числе геномная аннотация в формате GenBank. Как и в случае **FunGAP**, **funannotate** реализован на интерпретируемом языке Python 2.7, что ограничивает возможности его дальнейшей поддержки в случае обновление внутренних пакетов и их, как следствие, их несовместимости.

**LoReAn** (Cook et al., 2019) — это набор инструментов, который используетт транскриптомные прочтения, полученные с использованием технологий секвенирования Oxford Nanopore и Pacific Biosciences, для повышения качества предсказания белок-кодирующих генов. **LoReAn** обеспечивает повышенную точность аннотации, объединяя данные транскриптомного секвенирования с применением разных платформ секвенирования, референсные аминокислотные последовательности и результаты *ab initio* предсказания генов. Обработка данных происходит в два этапа:

Первый этап с небольшими модификациями соответствует инструменту **BAP** (Haas et al., 2011). В качестве дополнительных данных могут использоваться последовательности генов близких видов и видовое название референсной модели для инструмента **AUGUSTUS**. РНК-прочтения используются для предсказания генов при помощи **Braker** (Hoff et al., 2016), **AUGUSTUS** и **GeneMark-ES/ET**. Кроме того прочтения используются для сборки транскриптов при помощи **Trinity** (Grabherr et al., 2011). Полученные транскрипты выравниваются на геном при помощи **PASA** (Haas et al., 2008) и **GMAP** (Wu & Watanabe, 2005). **EVidenceModeler** (Haas et al., 2008) используется для объединения данных всех методов предсказания.

На втором этапе длинные прочтения картируют при помощи **GMAP**. Регионы, отвечающие экзонам определяются при помощи **gffread**, и, объединяя с данными первого этапа, кластеризуют и проводят реконструкцию транскриптов при помощи **bedtools** и **iAssembler**. Заключительную верификацию транскриптов проводят при помощи **GMAP** и **PASA**.

**Инструменты онлайн-аннотации**

**NCBI Eukaryotic Genome Annotation Pipeline** (Thibaud-Nissen et al., 2013) — это автоматизированный конвейер, который осуществляет аннотацию полных геномов и предварительных геномных сборок. Конвейер использует модульную структуру для выполнения всех задач аннотации, начиная с выборки необработанных и курируемых данных, с предпочтением последних, из общедоступных баз данных (**NCBI RefSeq**, **NCBI GenBank**, **SwissProt**, **TSA**,

**dbEST** и **NCBI SRA**), путем выравнивания последовательностей и предсказания генов, до конечной аннотации для публичных баз данных.

Основными компонентами конвейера являются программы выравнивания **Splign** (Kapustin et al., 2008) и **ProSplign** (Kiryutin et al., 2007) и программа предсказания генов **Gnomon** (Souvorov et al, 2010), объединяющая информацию из выравниваний экспериментальных данных и из моделей, созданных *ab initio* с помощью алгоритма на основе скрытых марковских моделей. Повторы предварительно маскируют при помощи **RepeatMasker** и **WindowMasker** (Morgulis et al., 2006). Аннотация РНК включает поиск микроРНК при помощи базы данных **miRBase** и **Splign**, тРНК при помощи **tRNAscan-SE**, рРНК и малых ядерных и ядрышковых РНК при помощи базы данных **Rfam** с использованием cmsearch из пакета **Infernal**. На середину 2020 года посредством NCBI EGAP было аннотировано 659 геномов.

**YGAP (**http://wolfe.ucd.ie/annotation/; Proux-Wéra E. et al. 2012) является онлайн инструментом аннотации геномных последовательностей дрожжей. **YGAP** использует существующие аннотированные геномы родственных видов дрожжей, для переноса структурных и функциональных аннотации генома с помощью **TBLASTN**. Аннотации генов тРНК выполняется при помощи **tRNAScan-SE** с использованием настроек по умолчанию. **YGAP** подходит для работы с последовательностями генома дрожжей прошедших и не прошедших полногеномную дупликацию. Входными данными является сборка генома в формате FASTA, содержащая не более 702 последовательностей, что исключает из анализа сильно фрагментированные предварительные сборки геномов, полученные из коротких прочтений. Дополнительно можно предоставить файл с прочтениями в формате FASTA, используемый для коррекции сдвигов рамки считывания, однако максимальный размер загружаемых файлов составляет 500 МБ, что ограничивает использование прочтений высокопроизводительного секвенирования, в частности полученных по технологии Illumina.

Помимо непосредственной аннотации **YGAP** генерирует несколько списком генов определенных категорий, которые позволяют судить о качестве генома: (1)

универсальные гены дрожжевых геномов (Gordon et al., 2009) не найденные в анализируемом геноме, (2) гены длиной более 150 а.к., не имеющие аннотированных гомологов, (3) значительно отличающие гомологи, определенные анализом синтении при помощи **SearchDOGS** (ÓhÉigeartaigh et al., 2011), (4) гены, не входящие в пространственные кластеры **YGOB** (Yeast Gene Order Browser): гены, уникальные для вида или рода, дуплицированные гены, гены, полученные в ходе горизонтального переноса генов и др.

Другим примером веб-платформы для структурной и функциональной аннотации генома является **GenSAS** (The **Gen**ome **S**equence **A**nnotation **S**erver; https://www.gensas.org; Humann et al., 2019), который объединяет геномные редакторы **JBrowse** (Ruels et al., 2016) и **Apollo** (Lee E. et al., 2013). Пользователи могут просматривать данные аннотаций и вручную создавать модели генов с использованием графического интерфейса, сопровождаемого подсказками и инструкциями. В качестве дополнительных данных пользователи могут загружать последовательности специфических для организма транскриптов и белков, а также транскриптомные данные для дальнейшего использования в процессе аннотирования. Проекты аннотирования геномов могут использоваться совместно с другими пользователями **GenSAS**, обеспечивая возможность совместной аннотации.

Основные этапы аннотации **GenSAS** включают: (1) идентификацию и маскирование повторов, (2) выравнивание доступных (публичных и подгруженных пользователем) данных на геном, (3) структурную аннотацию, (4) функциональную аннотацию белок-кодирующих генов, (5) дополнительное ручное редактирование генных моделей и создание окончательных файлов аннотаций. По завершении аннотации **GenSAS** генерирует файлы геномной аннотации в общераспространенных форматах.

**Инструменты оценки полноты геномных сборок**

Помимо непосредственно аннотации геномных последовательностей для исследователя важно понимать насколько качественной геномной сборкой он располагает. Помимо количественных метрик, таких как число контигов или скафолдов, N50, L50, N90, L90 важна информация о полноте и контаминации генома. Ниже мы рассмотрим несколько инструментов, позволяющих провести анализ полноты и контаминации генома.

**BUSCO** — **B**enchmarking sets of **U**niversal **S**ingle-**C**opy **O**rthologs (Seppey et al., 2019) — инструмент для оценки полноты генома по наборам маркерных генов из базы **OrthoDB** (https://www.orthodb.org/; Kriventseva et al, 2018). Программа может быть запущена в нескольких режимах: геномном, протеомном и транскриптомном в зависимости от типа входных данных (геномная последовательность, набор предсказанных белок-кодирующих генов и собранные транскрипты, соответственно). Пользователь может указать таксон, в рамках которого будет оцениваться качество генома или выполнить автоматических подбор подходящего таксона с использованием утилиты **SEPP** (https://github.com/smirarab/sepp). При указании таксона требуемые базы данных будут автоматически скачаны, если они не были загружены ранее. **BUSCO** присваивает каждому маркерному гену одно из четырёх состояний: полный и однокопийный, полный и многокопийный, фрагментированный или отсутствующий и выводит финальную статистику о качестве анализируемых данных.

**DOGMA** — это инструмент для оценки полноты протеома и транскриптома на основании последовательностей консервативных белковых доменов в предсказанных белок-кодирующих генах (https://domainworld.uni-muenster.de/programs/dogma/; Dohmen et al., 2016). Для оценки полноты генома используются наборы моделей из базы **Pfam** для следующих групп организмов: эукариоты, позвоночные, млекопитающие, членистоногие, насекомые, растения, двудольные (эвдикоты), однодольные (монокоты), грибы, бактерии, археи. Кроме того, пользователь может создавать собственные наборы генов. Для аннотирования доменов может применяться два инструмента на выбор: **Pfam_scan.pl** (скрипт на языке Perl), либо **RADIANT** (https://domainworld.uni-muenster.de/programs/radiant/), второй вариант позволяет ускорить обработку данных за счёт подгрузки базы в оперативную память (используется около 3 ГБ). Программа может быть использована онлайн [https://domainworld-services.uni-muenster.de/dogma/](https://domainworld-services.uni-muenster.de/dogma/), либо быть установлена локально. Исходный код написан на языке Python и доступен онлайн.

**FGMP** (Fungal Genome Mapping Pipeline) представляет собой инструмент для оценки полноты геномов грибов, реализованный на Perl (Cissé & Stajich, 2019). **FGMP** проводит анализ на основании 593 генов и 31 высококонсервативного геномного сегмента. Обработка данных происходит в три этапа:

1. Поиск однокопийных маркерных генов при помощи **EXONERATE**, инструментов **sixpack** и **csplit** из пакета **EMBOSS** (Rice et al., 2000) с последующим предсказанием белок-кодирующих генов при помощи **AUGUSTUS** и валидацией результата при помощи **pHMMER** (Eddy, 2011).
2. Определение представленности длинных консервативных некодирующих последовательностей ДНК, характерных для грибов при помощи **nHMMER** (Eddy, 2011).
3. Определение числа копий многокопийных белков при помощи **pHMMER** для оценки ошибок сборки, таких как объединение повторяющихся регионов в один.

Стоит отметить, что при помощи данного инструмента возможен анализ полноты генома на основании геномных прочтений, реализуемый при помощи **NCBI BLASTX**.

**Заключение**

На сегодняшний день большинство доступных инструментов аннотации эукариотических геномов обладают существенными недостатками, среди которых:

- неполный спектр реализуемых задач структурной и функциональной аннотации
- высокая специфичность для узкого спектра организмов

- необходимость предоставление специфических данных, например, протеомов близкородственных организмов или данных транскриптомного секвенирования
- особенности лицензирования, ограничивающие использование или доступ к исходному коду
- использование неактуальных и неподдерживаемых версий интерпретаторов языков программирования.
- сложность установки соответствующих версий всех необходимых зависимостей

По рассмотрении всего спектра программ стоит выделить два актуальных и свободно доступных инструмента: **BRAKER2** — программный пакет для комбинированного предсказания белок-кодирующих последовательностей и **Funannotate**, который обладает большой гибкостью в используемых входных данных и сочетает значительный набор инструментов для структурной и функциональной аннотации белок-кодирующих генов и некодирующих РНК.

**Конфликт интересов**

Авторы заявляют об отсутствии конфликта интересов.

**Conflict of interest**

The authors declare no conflict of interest.

**Список литературы / References**

Almagro Armenteros JJ, Tsirigos KD, Sønderby CK, et al. SignalP 5.0 improves signal peptide predictions using deep neural networks. Nature Biotechnology. 2019;37(4):420-423. doi:10.1038/s41587-019-0036-z

Altenhoff AM, Glover NM, Train C-M, et al. The OMA orthology database in 2018: retrieving evolutionary relationships among all domains of life through richer web and programmatic interfaces. Nucleic Acids Research. 2018;46(D1):D477-D485. doi:10.1093/nar/gkx1019

An J, Lai J, Lehman ML, Nelson CC. miRDeep*: an integrated application tool for miRNA identification from RNA sequencing data. Nucleic Acids Research. 2013;41(2):727-737. doi:10.1093/nar/gks1187

Arkin AP, Cottingham RW, Henry CS, et al. KBase: The United States Department of Energy Systems Biology Knowledgebase. Nature Biotechnology. 2018;36(7):566-569. doi:10.1038/nbt.4163

Ashburner M, Ball CA, Blake JA, et al. Gene Ontology: tool for the unification of biology. Nat Genet. 2000;25(1):25-29. doi:10.1038/75556

Bao W, Kojima KK, Kohany O. Repbase Update, a database of repetitive elements in eukaryotic genomes. Mobile DNA. 2015;6(1):11. doi:10.1186/s13100-015-0041-9

Behr J, Bohnert R, Zeller G, Schweikert G, Hartmann L, Rätsch G. Next generation genome annotation with mGene.ngs. BMC Bioinformatics. 2010;11(10):O8. doi:10.1186/1471-2105-11-S10-O8

Birney E, Clamp M, Durbin R. GeneWise and Genomewise. Genome Res. 2004;14(5):988-995. doi:10.1101/gr.1865504

Blin K, Shaw S, Steinke K, et al. antiSMASH 5.0: updates to the secondary metabolite genome mining pipeline. Nucleic Acids Research. 2019;47(W1):W81-W87. doi:10.1093/nar/gkz310

Brůna T, Hoff KJ, Lomsadze A, Stanke M, Borodovsky M. BRAKER2: automatic eukaryotic genome annotation with GeneMark-EP+ and AUGUSTUS supported by a protein database. NAR Genomics and Bioinformatics. 2021;3(lqaa108). doi:10.1093/nargab/lqaa108

Brůna T, Lomsadze A, Borodovsky M. GeneMark-EP+: eukaryotic gene prediction with self-training in the space of genes and proteins. NAR Genomics and Bioinformatics. 2020;2(lqaa026). doi:10.1093/nargab/lqaa026

Buchfink B, Xie C, Huson DH. Fast and sensitive protein alignment using DIAMOND. Nat Methods. 2015;12(1):59-60. doi:10.1038/nmeth.3176

Buels R, Yao E, Diesh CM, et al. JBrowse: a dynamic web platform for genome visualization and analysis. Genome Biology. 2016;17(1):66. doi:10.1186/s13059-016-0924-1

Bushmanova E, Antipov D, Lapidus A, Prjibelski AD. rnaSPAdes: a de novo transcriptome assembler and its application to RNA-Seq data. GigaScience. 2019;8(giz100). doi:10.1093/gigascience/giz100

Camacho C, Coulouris G, Avagyan V, et al. BLAST+: architecture and applications. BMC Bioinformatics. 2009;10(1):421. doi:10.1186/1471-2105-10-421

Cantarel BL, Korf I, Robb SMC, et al. MAKER: An easy-to-use annotation pipeline designed for emerging model organism genomes. Genome Res. 2008;18(1):188-196. doi:10.1101/gr.6743907

Chan PP, Lowe TM. tRNAscan-SE: Searching for tRNA Genes in Genomic Sequences. In: Kollmar M, ed. Gene Prediction: Methods and Protocols. Methods in Molecular Biology. Springer; 2019:1-14. doi:10.1007/978-1-4939-9173-0_1

Chang Z, Li G, Liu J, et al. Bridger: a new framework for de novo transcriptome assembly using RNA-seq data. Genome Biology. 2015;16(1):30. doi:10.1186/s13059-015-0596-2

Cissé OH, Stajich JE. FGMP: assessing fungal genome completeness. BMC Bioinformatics. 2019;20(1):184. doi:10.1186/s12859-019-2782-9

Cook DE, Valle-Inclan JE, Pajoro A, Rovenich H, Thomma BPHJ, Faino L. Long-Read Annotation: Automated Eukaryotic Genome Annotation Based on Long-Read cDNA Sequencing. Plant Physiology. 2019;179(1):38-54. doi:10.1104/pp.18.00848

Dobin A, Davis CA, Schlesinger F, et al. STAR: ultrafast universal RNA-seq aligner. Bioinformatics. 2013;29(1):15-21. doi:10.1093/bioinformatics/bts635

Dohmen E, Kremer LPM, Bornberg-Bauer E, Kemena C. DOGMA: domain-based transcriptome and proteome quality assessment. Bioinformatics. 2016;32(17):2577-2581. doi:10.1093/bioinformatics/btw231

Eddy SR. Accelerated Profile HMM Searches. Pearson WR, ed. PLoS Comput Biol. 2011;7(10):e1002195. doi:10.1371/journal.pcbi.1002195

El-Gebali S, Mistry J, Bateman A, et al. The Pfam protein families database in 2019. Nucleic Acids Research. 2019;47(D1):D427-D432. doi:10.1093/nar/gky995

Ellinghaus D, Kurtz S, Willhoeft U. LTRharvest, an efficient and flexible software for de novo detection of LTR retrotransposons. BMC Bioinformatics. 2008;9(1):18. doi:10.1186/1471-2105-9-18

Emms DM, Kelly S. OrthoFinder: solving fundamental biases in whole genome comparisons dramatically improves orthogroup inference accuracy. Genome Biology. 2015;16(1):157. doi:10.1186/s13059-015-0721-2

Finn RD, Attwood TK, Babbitt PC, et al. InterPro in 2017—beyond protein family and domain annotations. Nucleic Acids Research. 2017;45(D1):D190-D199. doi:10.1093/nar/gkw1107

Flynn JM, Hubley R, Goubert C, et al. RepeatModeler2 for automated genomic discovery of transposable element families. PNAS. 2020;117(17):9451-9457. doi:10.1073/pnas.1921046117

Galens K, Orvis J, Daugherty S, et al. The IGS Standard Operating Procedure for Automated Prokaryotic Annotation. Stand Genomic Sci. 2011;4(2):244-251. doi:10.4056/sigs.1223234

Gautheret D, Lambert A. Direct RNA motif definition and identification from multiple sequence alignments using secondary structure profiles11Edited by J. Doudna. Journal of Molecular Biology. 2001;313(5):1003-1011. doi:10.1006/jmbi.2001.5102

Gertz EM, Yu Y-K, Agarwala R, Schäffer AA, Altschul SF. Composition-based statistics and translated nucleotide searches: Improving the TBLASTN module of BLAST. BMC Biology. 2006;4(1):41. doi:10.1186/1741-7007-4-41

Girgis HZ. Red: an intelligent, rapid, accurate tool for detecting repeats de-novo on the genomic scale. BMC Bioinformatics. 2015;16(1):227. doi:10.1186/s12859-015-0654-5

Gordon JL, Byrne KP, Wolfe KH. Additions, Losses, and Rearrangements on the Evolutionary Route from a Reconstructed Ancestor to the Modern Saccharomyces cerevisiae Genome. PLOS Genetics. 2009;5(5):e1000485. doi:10.1371/journal.pgen.1000485

Gotoh O. Direct mapping and alignment of protein sequences onto genomic sequence. Bioinformatics. 2008;24(21):2438-2444. doi:10.1093/bioinformatics/btn460

Haas BJ, Delcher AL, Mount SM, et al. Improving the Arabidopsis genome annotation using maximal transcript alignment assemblies. Nucleic Acids Research. 2003;31(19):5654-5666. doi:10.1093/nar/gkg770

Haas BJ, Salzberg SL, Zhu W, et al. Automated eukaryotic gene structure annotation using EVidenceModeler and the Program to Assemble Spliced Alignments. Genome Biology. 2008;9(1):R7. doi:10.1186/gb-2008-9-1-r7

Haas BJ, Zeng Q, Pearson MD, Cuomo CA, Wortman JR. Approaches to Fungal Genome Annotation. Mycology. 2011;2(3):118-141. doi:10.1080/21501203.2011.606851

Hackenberg M, Rodríguez-Ezpeleta N, Aransay AM. miRanalyzer: an update on the detection and analysis of microRNAs in high-throughput sequencing experiments. Nucleic Acids Research. 2011;39(suppl_2):W132-W138. doi:10.1093/nar/gkr247

Haridas S, Salamov A, Grigoriev IV. Fungal Genome Annotation. In: de Vries RP, Tsang A, Grigoriev IV, eds. Fungal Genomics: Methods and Protocols. Methods in Molecular Biology. Springer; 2018:171-184. doi:10.1007/978-1-4939-7804-5_15

Hoff KJ, Lange S, Lomsadze A, Borodovsky M, Stanke M. BRAKER1: Unsupervised RNA-Seq-Based Genome Annotation with GeneMark-ET and AUGUSTUS. Bioinformatics. 2016;32(5):767-769. doi:10.1093/bioinformatics/btv661

Hölzer M, Marz M. De novo transcriptome assembly: A comprehensive cross-species comparison of short-read RNA-Seq assemblers. GigaScience. 2019;8(giz039). doi:10.1093/gigascience/giz039

Hubley R, Finn RD, Clements J, et al. The Dfam database of repetitive DNA families. Nucleic Acids Research. 2016;44(D1):D81-D89. doi:10.1093/nar/gkv1272

Huerta-Cepas J, Forslund K, Coelho LP, et al. Fast Genome-Wide Functional Annotation through Orthology Assignment by eggNOG-Mapper. Molecular Biology and Evolution. 2017;34(8):2115-2122. doi:10.1093/molbev/msx148

Humann JL, Lee T, Ficklin S, Main D. Structural and Functional Annotation of Eukaryotic Genomes with GenSAS. In: Kollmar M, ed. Gene Prediction: Methods and Protocols. Methods in Molecular Biology. Springer; 2019:29-51. doi:10.1007/978-1-4939-9173-0_3

Huntemann M, Ivanova NN, Mavromatis K, et al. The standard operating procedure of the DOE-JGI Microbial Genome Annotation Pipeline (MGAP v.4). Stand in Genomic Sci. 2015;10(1):86. doi:10.1186/s40793-015-0077-y

Palmer JM, Stajich J. Funannotate v1.8.1: Eukaryotic Genome Annotation. Zenodo; 2020. doi:10.5281/zenodo.4054262

Kalvari I, Argasinska J, Quinones-Olvera N, et al. Rfam 13.0: shifting to a genome-centric resource for non-coding RNA families. Nucleic Acids Research. 2018;46(D1):D335-D342. doi:10.1093/nar/gkx1038

Kanehisa M, Furumichi M, Tanabe M, Sato Y, Morishima K. KEGG: new perspectives on genomes, pathways, diseases and drugs. Nucleic Acids Res. 2017;45(D1):D353-D361. doi:10.1093/nar/gkw1092

Kanehisa M, Sato Y. KEGG Mapper for inferring cellular functions from protein sequences. Protein Sci. 2020;29(1):28-35. doi:10.1002/pro.3711

Kapustin Y, Souvorov A, Tatusova T, Lipman D. Splign: algorithms for computing spliced alignments with identification of paralogs. Biology Direct. 2008;3(1):20. doi:10.1186/1745-6150-3-20

Keilwagen J, Hartung F, Grau J. GeMoMa: Homology-Based Gene Prediction Utilizing Intron Position Conservation and RNA-seq Data. In: Kollmar M, ed. Gene Prediction: Methods and Protocols. Methods in Molecular Biology. Springer; 2019:161-177. doi:10.1007/978-1-4939-9173-0_9

Keilwagen J, Wenk M, Erickson JL, Schattat MH, Grau J, Hartung F. Using intron position conservation for homology-based gene prediction. Nucleic Acids Research. 2016;44(9):e89-e89. doi:10.1093/nar/gkw092

Keller O, Kollmar M, Stanke M, Waack S. A novel hybrid gene prediction method employing protein multiple sequence alignments. Bioinformatics. 2011;27(6):757-763. doi:10.1093/bioinformatics/btr010

Kent WJ. BLAT---The BLAST-Like Alignment Tool. Genome Research. 2002;12(4):656-664. doi:10.1101/gr.229202

Kiryutin B, Souvorov A, Tatusova T. ProSplign – Protein to Genomic Alignment Tool. In: Proc. 11th Annual International Conference in Research in Computational Molecular Biology. ; 2007.

Koonin EV, Fedorova ND, Jackson JD, et al. A comprehensive evolutionary classification of proteins encoded in complete eukaryotic genomes. Genome Biol. 2004;5(2):R7. doi:10.1186/gb-2004-5-2-r7

Korf I. Gene finding in novel genomes. BMC Bioinformatics. 2004;5(1):59. doi:10.1186/1471-2105-5-59

Kriventseva EV, Kuznetsov D, Tegenfeldt F, et al. OrthoDB v10: sampling the diversity of animal, plant, fungal, protist, bacterial and viral genomes for evolutionary and functional annotations of orthologs. Nucleic Acids Research. 2019;47(D1):D807-D811. doi:10.1093/nar/gky1053

Krogh A, Larsson B, von Heijne G, Sonnhammer ELL. Predicting transmembrane protein topology with a hidden markov model: application to complete genomes11Edited by F. Cohen. Journal of Molecular Biology. 2001;305(3):567-580. doi:10.1006/jmbi.2000.4315

Langmead B, Salzberg SL. Fast gapped-read alignment with Bowtie 2. Nature Methods. 2012;9(4):357-359. doi:10.1038/nmeth.1923

Laslett D, Canback B. ARAGORN, a program to detect tRNA genes and tmRNA genes in nucleotide sequences. Nucleic Acids Research. 2004;32(1):11-16. doi:10.1093/nar/gkh152

Laslett D, Canbäck B. ARWEN: a program to detect tRNA genes in metazoan mitochondrial nucleotide sequences. Bioinformatics. 2008;24(2):172-175. doi:10.1093/bioinformatics/btm573

Lee E, Helt GA, Reese JT, et al. Web Apollo: a web-based genomic annotation editing platform. Genome Biology. 2013;14(8):R93. doi:10.1186/gb-2013-14-8-r93

Li L, Stoeckert CJ, Roos DS. OrthoMCL: Identification of Ortholog Groups for Eukaryotic Genomes. Genome Research. 2003;13(9):2178-2189. doi:10.1101/gr.1224503

Liu J, Xiao H, Huang S, Li F. OMIGA: Optimized Maker-Based Insect Genome Annotation. Mol Genet Genomics. 2014;289(4):567-573. doi:10.1007/s00438-014-0831-7

Lombard V, Golaconda Ramulu H, Drula E, Coutinho PM, Henrissat B. The carbohydrate-active enzymes database (CAZy) in 2013. Nucleic Acids Research. 2014;42(D1):D490-D495. doi:10.1093/nar/gkt1178

Lomsadze A, Burns PD, Borodovsky M. Integration of mapped RNA-Seq reads into automatic training of eukaryotic gene finding algorithm. Nucleic Acids Research. 2014;42(15):e119-e119. doi:10.1093/nar/gku557

Lowe TM, Eddy SR. tRNAscan-SE: A Program for Improved Detection of Transfer RNA Genes in Genomic Sequence. Nucleic Acids Research. 1997;25(5):955-964. doi:10.1093/nar/25.5.955

Majoros WH, Pertea M, Salzberg SL. TigrScan and GlimmerHMM: two open source ab initio eukaryotic gene-finders. Bioinformatics. 2004;20(16):2878-2879. doi:10.1093/bioinformatics/bth315

Min B, Grigoriev IV, Choi I-G. FunGAP: Fungal Genome Annotation Pipeline using evidence-based gene model evaluation. Bioinformatics. 2017;33(18):2936-2937. doi:10.1093/bioinformatics/btx353

Morgulis A, Gertz EM, Schäffer AA, Agarwala R. WindowMasker: window-based masker for sequenced genomes. Bioinformatics. 2006;22(2):134-141. doi:10.1093/bioinformatics/bti774

Nawrocki EP, Eddy SR. Infernal 1.1: 100-fold faster RNA homology searches. Bioinformatics. 2013;29(22):2933-2935. doi:10.1093/bioinformatics/btt509

ÓhÉigeartaigh SS, Armisén D, Byrne KP, Wolfe KH. Systematic discovery of unannotated genes in 11 yeast species using a database of orthologous genomic segments. BMC Genomics. 2011;12(1):377. doi:10.1186/1471-2164-12-377

Proux-Wéra E, Armisén D, Byrne KP, Wolfe KH. A pipeline for automated annotation of yeast genome sequences by a conserved-synteny approach. BMC Bioinformatics. 2012;13(1):237. doi:10.1186/1471-2105-13-237

Quevillon E, Silventoinen V, Pillai S, et al. InterProScan: protein domains identifier. Nucleic Acids Res. 2005;33(Web Server issue):W116-120. doi:10.1093/nar/gki442

Rawlings ND, Barrett AJ, Bateman A. MEROPS: the peptidase database. Nucleic Acids Res. 2010;38(Database issue):D227-233. doi:10.1093/nar/gkp971

Rice P, Longden I, Bleasby A. EMBOSS: the European Molecular Biology Open Software Suite. Trends Genet. 2000;16(6):276-277. doi:10.1016/s0168-9525(00)02024-2

Robertson G, Schein J, Chiu R, et al. De novo assembly and analysis of RNA-seq data. Nature Methods. 2010;7(11):909-912. doi:10.1038/nmeth.1517

Sallet E, Gouzy J, Schiex T. EuGene: An Automated Integrative Gene Finder for Eukaryotes and Prokaryotes. In: Kollmar M, ed. Gene Prediction: Methods and Protocols. Methods in Molecular Biology. Springer; 2019:97-120. doi:10.1007/978-1-4939-9173-0_6

Sallet E, Gouzy J, Schiex T. EuGene-PP: a next-generation automated annotation pipeline for prokaryotic genomes. Bioinformatics. 2014;30(18):2659-2661. doi:10.1093/bioinformatics/btu366

Seppey M, Manni M, Zdobnov EM. BUSCO: Assessing Genome Assembly and Annotation Completeness. In: Kollmar M, ed. Gene Prediction: Methods and Protocols. Methods in Molecular Biology. Springer; 2019:227-245. doi:10.1007/978-1-4939-9173-0_14

Slater GSC, Birney E. Automated generation of heuristics for biological sequence comparison. BMC Bioinformatics. 2005;6(1):31. doi:10.1186/1471-2105-6-31

Souvorov A, Kapustin Y, Kiryutin B, Chetvernin V., Tatusova T., Lipman D. Gnomon–NCBI eukaryotic gene prediction tool. In: National Center for Biotechnology Information. (2010) https://www.ncbi.nlm.nih.gov/core/assets/genome/files/Gnomon-description.pdf

Stanke M, Diekhans M, Baertsch R, Haussler D. Using native and syntenically mapped cDNA alignments to improve de novo gene finding. Bioinformatics. 2008;24(5):637-644. doi:10.1093/bioinformatics/btn013

Tanizawa Y, Fujisawa T, Nakamura Y. DFAST: a flexible prokaryotic genome annotation pipeline for faster genome publication. Hancock J, ed. Bioinformatics. 2018;34(6):1037-1039. doi:10.1093/bioinformatics/btx713

Tempel S. Using and Understanding RepeatMasker. In: Bigot Y, ed. Mobile Genetic Elements: Protocols and Genomic Applications. Methods in Molecular Biology. Humana Press; 2012:29-51. doi:10.1007/978-1-61779-603-6_2

Ter-Hovhannisyan V, Lomsadze A, Chernoff YO, Borodovsky M. Gene prediction in novel fungal genomes using an ab initio algorithm with unsupervised training. Genome Res. 2008;18(12):1979-1990. doi:10.1101/gr.081612.108

Testa AC, Hane JK, Ellwood SR, Oliver RP. CodingQuarry: highly accurate hidden Markov model gene prediction in fungal genomes using RNA-seq transcripts. BMC Genomics. 2015;16(1):170. doi:10.1186/s12864-015-1344-4

Thibaud-Nissen F, Souvorov A, Murphy T, DiCuccio M, Kitts P. Eukaryotic genome annotation pipeline. In The NCBI Handbook [Internet]. 2nd edition. National Center for Biotechnology Information (US). 2013

Thorvaldsdóttir H, Robinson JT, Mesirov JP. Integrative Genomics Viewer (IGV): high-performance genomics data visualization and exploration. Briefings in Bioinformatics. 2013;14(2):178-192. doi:10.1093/bib/bbs017

Trapnell C, Roberts A, Goff L, et al. Differential gene and transcript expression analysis of RNA-seq experiments with TopHat and Cufflinks. Nature Protocols. 2012;7(3):562-578. doi:10.1038/nprot.2012.016

Wang X, Zhang J, Li F, et al. MicroRNA identification based on sequence and structure alignment. Bioinformatics. 2005;21(18):3610-3614. doi:10.1093/bioinformatics/bti562

Wu TD, Watanabe CK. GMAP: a genomic mapping and alignment program for mRNA and EST sequences. Bioinformatics. 2005;21(9):1859-1875. doi:10.1093/bioinformatics/bti310

Xie Y, Wu G, Tang J, et al. SOAPdenovo-Trans: de novo transcriptome assembly with short RNA-Seq reads. Bioinformatics. 2014;30(12):1660-1666. doi:10.1093/bioinformatics/btu077

Yandell M, Ence D. A beginner's guide to eukaryotic genome annotation. Nat Rev Genet. 2012;13(5):329-342. doi:10.1038/nrg3174

Yang J-H, Zhang X-C, Huang Z-P, et al. snoSeeker: an advanced computational package for screening of guide and orphan snoRNA genes in the human genome. Nucleic Acids Research. 2006;34(18):5112-5123. doi:10.1093/nar/gkl672

Zhang S, Li D, Zhang G, Wang J, Niu N. The Prediction of Rice Gene by Fgenesh. Agricultural Sciences in China. 2008;7(4):387-394. doi:10.1016/S1671-2927(08)60081-4